\documentclass{article}
\usepackage[accepted]{icml2022}

\usepackage{times}
\usepackage{soul}
\usepackage{url}
\usepackage[utf8]{inputenc}
\usepackage[small]{caption}
\usepackage{graphicx}
\usepackage{amsmath}
\usepackage{amsthm}
\usepackage{booktabs}
\usepackage{algorithm}
\usepackage{algorithmic}
\usepackage{amsmath}
\usepackage{amsfonts,amssymb}
\usepackage{bm}
\usepackage{multirow}
\usepackage{bbm}
\usepackage{color}
\usepackage{subfigure}

\newcommand{\cv}[0]{\ensuremath{\boldsymbol{c}} }

\newcommand{\rv}[0]{\ensuremath{\boldsymbol{r}} }

\newcommand{\uv}[0]{\ensuremath{\boldsymbol{u}} }

\newcommand{\vv}[0]{\ensuremath{\boldsymbol{v}} }
\newcommand{\Av}[0]{\ensuremath{\boldsymbol{A}} }
\newcommand{\Bv}[0]{\ensuremath{\boldsymbol{B}} }

\newcommand{\Yv}[0]{\ensuremath{\boldsymbol{Y}} }

\newcommand{\phiv}[0]{\ensuremath{\boldsymbol{\phi}} }
\newcommand{\Phiv}[0]{\ensuremath{\boldsymbol{\Phi}} }

\newcommand{\thetav}[0]{\ensuremath{\boldsymbol{\theta}} }
\newcommand{\Thetav}[0]{\ensuremath{\boldsymbol{\Theta}} }

\newcommand{\gammav}[0]{\ensuremath{\boldsymbol{\gamma}} }

\newcommand{\Omegav}[0]{\ensuremath{\boldsymbol{\Omega}} }

\begin{document}
	
	\twocolumn[
	\icmltitle{
		Ordinal Graph Gamma Belief Network for Social Recommender Systems
	}
	
	
	
	\icmlsetsymbol{equal}{*}

	\begin{icmlauthorlist}
		\icmlauthor{Dongsheng Wang}{1}
		\icmlauthor{Chaojie Wang}{1}
		\icmlauthor{Bo Chen}{1}
		\icmlauthor{Mingyuan Zhou}{2}
	\end{icmlauthorlist}
	
	\icmlaffiliation{1}{Xidian University}
	\icmlaffiliation{2}{The University of Texas at Austin}
	
	\icmlcorrespondingauthor{Bo Chen}{bchen@mail.xidian.edu.cn}
	
	\icmlkeywords{Machine Learning, ICML}
	
	\vskip 0.3in]

	
	
	\printAffiliationsAndNotice{}  

\begin{abstract}
To build recommender systems that not only consider user-item interactions represented as ordinal variables, but also exploit the social network describing the relationships between the users,
we develop a hierarchical Bayesian model termed ordinal graph factor analysis (OGFA), which jointly models user-item and user-user interactions.
OGFA not only achieves good recommendation performance, but also extracts interpretable latent factors corresponding to representative user preferences. We further extend OGFA to ordinal graph gamma belief network, which is a multi-stochastic-layer deep probabilistic model that captures the user preferences and social communities at multiple semantic levels.
For efficient inference, we develop a parallel hybrid Gibbs-EM algorithm, which exploits the sparsity of the graphs and is scalable to large datasets. Our experimental results show that the proposed models not only outperform recent baselines on recommendation datasets with explicit or implicit feedback, but also provide interpretable latent representations.
\end{abstract}

\vspace{-3mm}
\section{Introduction}
	There is a surge
	of research interest in constructing recommender systems based on the observed user-item interactions.
	Collaborative filtering (CF) \cite{bennett2007netflix} is a popular recommendation technique that has achieved state-of-the-art performance, which is often barely based on the users' feedback on items, including explicit feedback such as ratings, or implicit one such as quantized play counts \cite{sarwar2001item,hu2008collaborative}. The feedback of $U$ users on $I$ items can often be represented as a $U \times I$ matrix $ \Yv $ of ordinal variables,  which are categorical data exhibiting natural ordering between categories (for example: $unlike \prec normal \prec like$).
	Such data are known to be sparse, bursty, and over-dispersed, making their direct use in recommender systems challenging.

	Matrix Factorization (MF) \cite{koren2009matrix} is a popular approach of CF algorithms, 
	which aims to approximate the observations by a low-rank structure:
	$\Yv \approx \Thetav \Phiv^T$, where $\Thetav$ and $\Phiv$ describe user preferences and item attributes, respectively.
	Thus each user or item can be represented as a $K$-dimensional latent vector, where $K \ll \mbox{min}(U,I)$, and the user-item interaction can be measured by their corresponding inner product. Among these methods based on MF, Poisson factorization (PF) \cite{gopalan2015scalable} is well suited for count data, which replaces the usual Gaussian assumption with a Poisson one, and has become popular in handling implicit feedback, achieving SOTA results. 
	PF is also often applied to a binarized version of the user-item interaction matrix, containing only the information that a user is interacting with an item or not.
	However, the binarization stage induces a loss of information for PF, since the value associated to each interaction is removed.
	Although there have been several attempts in the literature to directly model raw
	data with PF \cite{gopalan2015scalable}, or introduce the Bernoulli-Poisson link for binarized data, referred to as Bernoulli-Poisson factorization (BePoF) \cite{acharya2015nonparametric}, these PF based methods still fail to fully describe the ordinal nature of the ordinal data within 
	a limited range.
	Rather than treating the ordinal feedback as binary or count variables,
	several recent works have tried directly modeling the raw ordinal data.
	Discrete compound Poisson factorization (dcPF) \cite{gouvert2020recommendation} adds a latent variable that obeys exponential dispersion model (EDM) in its generative process.
	\cite{agresti2003categorical} utilize Cumulative Link Models (CLMs) to make a bridge between ordinal data and MF models, leading to OrdMF models \cite{gouvert2020ordinal}.
	The success of OrdMF motivates us to construct a fully Bayesian model to directly handle raw ordinal data.

	From another perspective, moving beyond traditional recommender systems where the users are treated as independent and identically distributed ($i.i.d.$), many researchers have recently started to analyze social recommender systems
	\cite{tang2013social,jiang2014scalable}
	meeting the prevalence of online social networks.
	Such social recommendation approaches are based on the social influence theory that states connected people would influence each other, leading to shared interests due to social interactions \cite{anagnostopoulos2008influence,lewis2012social}.
	For instance, social regularization (or graph regularization) has been empirically proven effective for social recommendation, by assuming that connected users would share similar latent embeddings \cite{cai2010graph,chang2009relational,acharya2015gamma}.

	
	%
	%
	

    In an attempt to keep as much ordinal data information as possible and also consider social networks, we first develop a novel probabilistic generative model for jointly modeling both ordinal user-item interactions and user-user relations, named ordinal graph factor analysis (OGFA).
	Further, we extend OGFA to a hierarchical fashion to discover both the underlying user preferences and social communities (or groups) at different semantic levels.
	The contributions of this paper are listed as follows:
${1)}$ We propose 
			OGFA to jointly model both the raw ordinal user-item interactions and user-user social network via sharing their latent representations;
${2)}$ We extend OGFA to a hierarchical version, named Ordinal Graph Gamma Belief Network (OG-GBN), to provide multi-layer user latent representations, revealing both their preferences and relationships at different semantic levels;
${3)}$ We integrate high-order social information into the deep structure of OG-GBN to model the recursive dynamic social diffusion, which is a common problem in social recommendation;
${4)}$ For efficient inference, we develop a  parallel hybrid Gibbs-EM algorithm for our models. It can make full use of the sparsity of the observed matrix and is scalable to large datasets.
			
\vspace{-1mm}
	\section{Related Work}

	\noindent\textbf{Recommendation with Raw Ordinal Data:   }
	Neither PFA nor BerPo-PFA
	make correct assumptions when modeling
	ordinal data. One may better handle overdispersion by factorizing the rating matrix under the negative binomial (NB) likelihood like NBFA \cite{zhou2018nonparametric}, but this choice still ignores the fact that ordinal data take values from a limited range.  
	Many efforts have been devoted to developing generative processes using the raw ordinal data to achieve better representation and recommendation performance.
	Discrete compound Poisson factorization (dcPF) \cite{gouvert2020recommendation} introduces a discrete exponential dispersion models (EDM) as a map function, which links the latent count variable to the discrete observation. Inspired by the same ideals, Ordinal NMF (OrdNMF) \cite{gouvert2020ordinal} methods adapt the
	Cumulative Link Models (CLMs) have been applied for ordinal regression \cite{agresti2003categorical} and the underlying thought is to introduce a step function. 

	\noindent\textbf{Social Recommendation Systems:   }
	Many researchers have recently focused on social recommender systems \cite{tang2013social,jiang2014scalable,fan2019graph}, which has been emerging as a promising direction of analyzing user preferences via incorporating social information \cite{guo2012simple}.
	%
	%
	For instances,
	considering the influence of trust users (including both trustees and trusters) on the rating prediction, \cite{guo2015trustsvd} develop a TrustSVD based on SVD++ \cite{koren2008factorization} and ensure that user-specific vectors can be learned from their trust information even if a few or no rating are given;
	%
	%
	Relational topic models (RTMs) treat each document as a user and construct a probabilistic model to jointly model observed document-word (user-item) matrix and document-document (user-user) interactions 
	\cite{chang2009relational,rosen2012author,acharya2015gamma}.
	We argue that 1) RTMs simply unitize Poisson likelihood to model the rating matrix which has limited representation of ordinal data \cite{gouvert2020recommendation}, and
	2) the above social networks only employ the first-order social information, while ignoring social diffusion when making a recommendation.

\vspace{-1mm}
\section{Preliminary}
    For comprehensively understanding the importance of modeling raw ordinal data, we introduce the following background.

	\noindent\textbf{Poisson Factor Analysis: } \label{sub_mf}
 PFA \cite{gopalan2015scalable} is a typical topic model and serves as a building block for our developed models.
	A common preprocessing operation for applying PFA (or other sophisticated variants) to recommendation is to binarize the observed user-item interaction matrix $ \Yv \in \mathbb{Z}^{U \times I}, \mathbb{Z}:=\{0,1,...\}$ to a binary one $\Bv \in {\{ 0, 1 \}}^{U \times I}$, where $U$ and $I$ indicate the number of user and item, respectively.
	Then the binarized user-item interaction matrix $\Bv$ can be factorized into a summation of $K$ equal-size latent matrices under the Poisson likelihood, formulated as
	\begin{equation} \label{eq_pfa}
	\setlength{\abovedisplayskip}{3pt}
    \setlength{\belowdisplayskip}{3pt}
	\Bv \sim \mbox{Pois}(\Thetav \Phiv^T),
	\end{equation}
	where $\Thetav \in \mathbb{R}_+^{U \times K}$ and $\Phiv \in \mathbb{R}_+^{I \times K}$ denote the factor loading matrix and factor scores, respectively, with $\mathbb{R}_+=\{x:x\ge0\}$.
	More specifically, each row of $\Thetav$, denoted as $\thetav_u \in \mathbb{R}_+^{K}$, contains relative community intensities specific to user $u$;
each column of $\Phiv$, denoted as $\phiv_k \in \mathbb{R}_+^{I}$, encodes the relative importance of each item in  community $k$.
    Thus each binarized observation $b_{ui}$ can be modeled with a Poisson distribution and further parameterized by the inner product of corresponding user preference and item attributes as
    \begin{equation}
    \setlength{\abovedisplayskip}{3pt}
    \setlength{\belowdisplayskip}{3pt}
    b_{ui} \sim \mbox{Poisson}(\thetav_{u}^{T}\phiv_{i}),
    \end{equation}
    where the sparsity of gamma distributed $\thetav_u$ indicates that each user is only interested in a few communities, contributing to exploring both the underlying user preferences and social communities.

	\vspace{1mm}
	
	\noindent\textbf{Bernoulli-Poisson Link: }
	Instead of directly factorizing the binarized matrix $\Bv$ under the Poisson likelihood, a better solution could be linking these binary interactions to latent count values with the Bernoulli-Poisson (BerPo) link \cite{zhou2015infinite} and then factorizing the latent count matrix.
	Below we consider modeling user-item interactions after binarization with BerPo-PFA, where $b_{ui}:=\Bv(u,i)$ equals to one if and only if nodes $u$ and $i$ are linked. Thus each nonzero interaction $b_{ui}$ can be assumed to be derived as:
	\begin{equation}
	\setlength{\abovedisplayskip}{3pt}
    \setlength{\belowdisplayskip}{3pt}
	\small
	\textstyle b_{ui}=\mathbbm{1}(m_{ui} \geq 1),\ m_{ui} \sim \mbox{Pois}(\lambda_{ui}),\ \lambda_{ui} = \sum_{k} \theta_{uk} \phi_{ik} \notag
	\end{equation}
	where $m_{ui} \in \mathbb{Z}$ denotes the augmented latent count variable,
	$\mathbbm{1}(\cdotp)$ an indicator function, and
	$\lambda_{ui}$ the inner product of the corresponding latent representations of user $u$ and item $i$, specifically $\thetav_u$ and $\phiv_i$.
	Integrating $m_{ui}$ out leads to the following generative process:
	$ 
	b_{ui} \sim \mbox{Bern}(1-e^{-\lambda_{ui}})\notag
	$, 
	where $\mbox{Bern}$ refers to the Bernoulli distribution. The conditional posterior of $m_{ui}$ can be expressed as
	\begin{equation} \label{eq_delta}
	\setlength{\abovedisplayskip}{3pt}
    \setlength{\belowdisplayskip}{3pt}
	(m_{ui}|b_{ui},\lambda_{ui}) \sim \left\{
	\begin{aligned}
	&\delta_0, &if\ b_{ui}=0,\\
	&\mbox{Pois}_{+}(\lambda_{ui}), &if\ b_{ui}=1,
	\end{aligned}
	\right.
	\end{equation}
	where $\mbox{Pois}_{+}$ refers to the zero-truncated Poisson distribution and $\delta_0$ to the Dirac distribution located in 0.
    Moving beyond user-item interactions, we note that the BerPo-PFA can be also directly applied to model the binarized user-user relations, denoted as $\Av \in {\{ 0, 1 \}}^{U \times U}$.

	\vspace{1mm}
	
	\noindent\textbf{Cumulative Link Model for Ordinal Data: }
	Moving beyond adopting the binarization stage, which discards the value information associated to the user-item interaction, Cumulative Link Model (CLM) is developed to mix up the gap between the raw ordinal data and the factorization models \cite{agresti2003categorical}, playing a role of threshold model for ordinal regression without loss of generality.
	More specifically, for each observed ordinal user-item interaction $y_{ui}:=\Yv(u,i)$, CLM introduces a continuous latent variable $x_{ui} \in \mathbb{R}_+:=[0,+\infty)$, which can be mapped to $y_{ui}$ with a number of contiguous intervals with boundaries $\{ {b_v}\} _{v = 0}^{V}$ as following:
	\begin{equation}\label{eq_sequence}
	\setlength{\abovedisplayskip}{3pt}
    \setlength{\belowdisplayskip}{3pt}
	0=b_0<...<b_{V-1}<b_V= +\infty.
	\end{equation}
Thus the rate score $\lambda_{ui}$ meeting the interval $[b_{v-1},b_{v})$ will be assigned to the corresponding rate $v$ as
	\begin{equation}
	\setlength{\abovedisplayskip}{3pt}
    \setlength{\belowdisplayskip}{3pt}
	\lambda_{ui} \longrightarrow y_{ui}=v, \ if \ \lambda_{ui} \in [b_{v-1},b_v). \notag
	\end{equation}
	The projection between $\lambda_{ui}$ and $y_{ui}$ can be defined with a quantization function as
	\begin{equation}\label{eq_G_x}
	\setlength{\abovedisplayskip}{3pt}
    \setlength{\belowdisplayskip}{3pt}
	y_{ui} = G_b({\lambda_{ui}}) \in \mathbb{Z}:=\{0,...,V\}. \notag
	\end{equation}
	%
	
	To ensure the non-genativity of $\lambda_{ui}$ and the ability to model over-dispersion, Ordinal NMF \cite{gouvert2020ordinal} typically introduces a non-negative multiplicative random noise $\varepsilon_{ui}$, with c.d.f denoted as $F_{\varepsilon}(\cdot)$, on the latent variable $\lambda_{ui}$ as
	$\ y_{ui} = G_b(\lambda_{ui}  \varepsilon_{ui})$.
	Thus the c.d.f. of the ordinal random variable $y_{ui}$ can be formulated as:
	\begin{align}
	\setlength{\abovedisplayskip}{3pt}
    \setlength{\belowdisplayskip}{3pt}
	p[y_{ui} \leq v |\lambda_{ui}] &= p[G_b(\lambda_{ui} \varepsilon_{ui}) \leq v | \lambda_{ui}]  = p[\lambda_{ui}  \varepsilon_{ui} < b_v]\notag\\
	&\textstyle = p[\varepsilon_{ui}  < \frac{b_v}{\lambda_{ui}}] = F_{\varepsilon}(\frac{b_v}{\lambda_{ui}}), \label{eq_cdf}
	\end{align}
	where the p.m.f. can be calculated with:
	\begin{align}\label{eq_pmf}
	\setlength{\abovedisplayskip}{3pt}
    \setlength{\belowdisplayskip}{3pt}
	p[y_{ui}=v|\lambda_{ui}]&= p[y_{ui} \leq v|\lambda_{ui}] - p[y_{ui} \leq v-1|\lambda_{ui}]\notag\\
	&\textstyle = F_{\varepsilon}(\frac{b_v}{\lambda_{ui}}) - F_{\varepsilon}(\frac{b_{v-1}}{\lambda_{ui}}).
	\end{align}
	Notably, various functions $F_{\varepsilon}(\cdot)$ can be used to determine the exact nature of the multiplicative noise.

\vspace{-1mm}
\section{Ordinal Graph Gamma Belief Network}
	
Focusing on constructing a probabilistic generative model that can not only directly handle with raw ordinal data, but also jointly model both  user-item and user-user interactions, we propose OGFA and further extend it to a hierarchical version, discovering representative user preferences and user communities at different semantic levels.


	\vspace{1mm}
	
\begin{figure}[t]
		\centering
		\includegraphics[width=50mm]{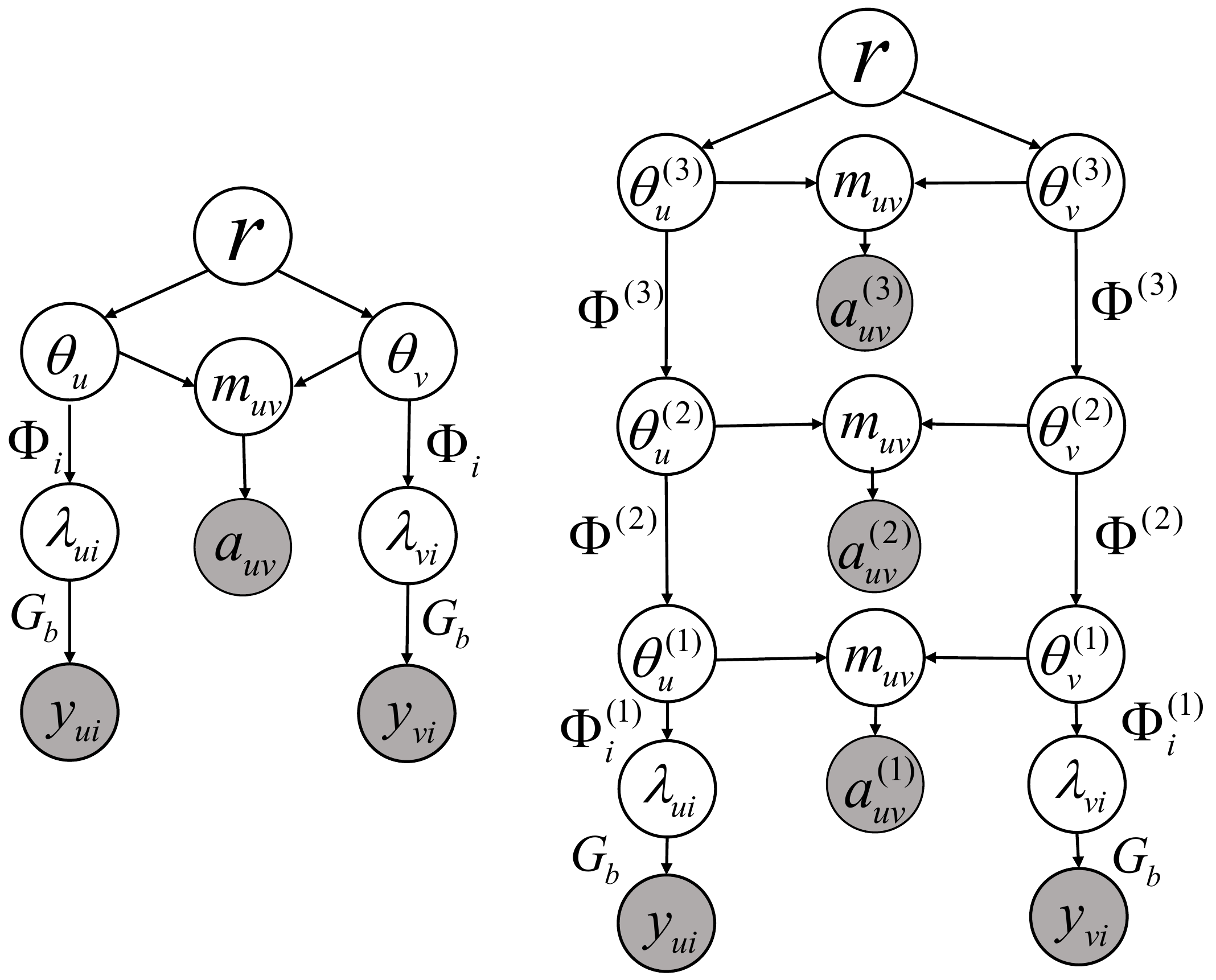}
		\caption{\small Graphical descriptions of OGFA (left) and a 3-layer OG-GBN (right).
		}
		\label{figure:relation}
		\vspace{-4mm}
	\end{figure}

	\noindent\textbf{Ordinal Graph Factor Analysis:   }
	%
	%
	To consistent with the definition in the preliminary, we denote the raw ordinal user-item matrix as
	$ \Yv \in \mathbb{Z}^{U \times I}$.
	Then the user-user social network can be represented as a set of users (nodes) and their relations (edges), resulting in a graph of size $U \times U$.
	In this paper, we focus on analyzing the undirected case, and represent the undirect graph as a symmetric binary adjacency matrix $\Av \in \{0,1\}^{U \times U}$,
which can be linked to a latent count matrix with the BerPo link in the preliminary.
	As illustrated  in the left part of Fig.~\ref{figure:relation}, the generative model of the proposed OGFA for jointly modeling user-item interaction $\Yv$ and user-user social network $\Av$ can be formulated as:
	\begin{equation} \label{OGFA}
	\begin{split}
	&a_{uv} = \mathbbm{1}(m_{uv} \geq 1),\ y_{ui} \sim \mbox{G}_b(\lambda_{ui}), \\
	&m_{uv} = \sum_k m_{ukv}, \ m_{ukv} \sim \mbox{Pois}(\theta_{uk} u_k \theta_{vk}),\\
	&{\lambda _{ui}} = \sum\limits_k {{\theta _{uk}}{\phi _{ik}}}, \ \phiv_k \sim \mbox{Dir}(\eta_k), \\
	&\thetav_{u} \sim \mbox{Gam}(\rv, 1/c_u),\ u_k \sim \mbox{Gam}(\gamma /K, 1/c_0)
	\end{split}
	\end{equation}
	where $\mbox{G}_b$ is the quantization function defined in Eq.~\eqref{eq_G_x};
	$m_{ukv}$ represents how often users $u$ and  $v$ interact due to their affiliations with community $k$;
	$u_k$, which measures the importance of community $k$ in explaining user-user interactions, also contributes to balance the scale of $\theta_{uk}$ that is related to both  
	$a_{uv}$ and $y_{ui}$.
	Notably, the $K$ latent communities in OGFA are treated independently and our model can be easily extended to the situation of considering both intra- and inter- community interactions, through modifying the generative process with respect to $m_{uv}$ as
	\begin{equation}\small\textstyle
	m_{uv} = \sum_{k_1} \sum_{k_2} m_{uvk_1k_2}, \ m_{uvk_1k_2} \sim \mbox{Pois}(\theta_{uk_1}u_{k_1k_2}\theta_{vk_2}).\notag
	\end{equation}
	In our consideration, treating each community independently will be more suitable to model an assortative relational network exhibiting homophily (e.g., co-author network) and similar conclusion can be found in \cite{zhou2015infinite}.
	Thus in what follows, we focus on the former simpler model by omitting inter-community interactions.
	
	The main purpose of OGFA is to jointly infer the user preferences $\thetav_u$ and item attributes $\phiv_i$ as well as the threshold sequence $\{ {b_v}\} _{v = 0}^{V}$ in Eq.~\eqref{eq_sequence}.
Focusing on the special case, where $\varepsilon_{ui}$ is a multiplicative inverse-gamma (IG) noise with the shape parameter $\alpha=1$, specifically  $\varepsilon_{ui} \sim \mbox{IG}(1,1)$ with the c.d.f $F_{\varepsilon}(x)=e^{-1/x}$, the c.d.f in Eq.~\eqref{eq_cdf} associated with the ordinal data $y_{ui}$ can be obtained as
	\begin{align}\label{eq_cdf_2}
	p(y_{ui} \leq v|\lambda_{ui}) &= e^{-\lambda_{ui}\gamma_v}, 
	\end{align}
	where $v \in \{0,...,V\}$ and $\gamma_v=b_v^{-1}$. The sequence $\gamma_{0:V}$ corresponds to the inverse of the thresholds and is therefore decreasing, i.e., $\gamma_{-1}=+\infty>\gamma_0>...>\gamma_{V-1}>\gamma_V=0$.
	Moreover, defining $\Delta_v:=\gamma_{v-1}-\gamma_v$ for $v \in \{1,...,V\}$, we have $\gamma_v=\sum_{l=v+1}^{V}\Delta_l$, specifically defining $\gamma_{V-1}=\Delta_V$. The event
	${\{ {y_{ui}}\} _{{y_{ui}} > v}}$
	satisfies a Bernoulli distribution as
	\begin{equation}
	p({y_{ui}} > v|{\lambda _{ui}}) = \mbox{Bern}(1-e^{-\lambda_{ui}\gamma_v}). \notag
	\end{equation}
	It is interesting to notice these BerPo-based relational models \cite{zhou2015infinite,acharya2015nonparametric} can be regarded as special cases of the OGFA setting $V=1$ and $\gamma_0=1$.
	
    Combining Eq.\eqref{eq_pmf} and \eqref{eq_cdf_2}, the p.m.f. of the observed $y_{ui}$ can be obtained as :
	\begin{equation}
	\begin{small}
	p(y_{ui}=v|\lambda_{ui})=\left\{
	\begin{aligned}
	&e^{-\lambda_{ui}\gamma_0}, &\ v=0,\\
	&e^{-\lambda_{ui}\gamma_v}-e^{\lambda_{ui}\gamma_{v-1}}, &\ 1\leq v <V,\\
	&1-e^{-\lambda_{ui}\gamma_{V-1}}, &\ v=V\\
	\end{aligned}
	\right. \notag
	\end{small}
	\end{equation}
	where the log-likelihood of $\lambda_{ui}$ can be formulated as:
	\begin{equation}
	\label{eq_log_like}
	\begin{small}
	\log p(y_{ui}=v|\lambda_{ui})=\left\{
	\begin{aligned}
	&-\lambda_{ui}\gamma_0, &v=0\\
	&-\lambda_{ui}\gamma_v + \log (1-e^{-\lambda_{ui}\Delta_v}), &v>0\\
	\end{aligned}
	\right.
	\end{small}
	\end{equation}
	Notably, the log-likelihood in Eq.~\eqref{eq_log_like} not only brings up a linear term in $\lambda_{ui}$ when $v=0$, but also a nonlinear term of the form $\lambda_{ui}$$\longrightarrow$$ \log(1-e^{-\lambda_{ui}})$, where the latter part is
	similar to the Bernoulli-Poisson link.
	
	\vspace{1mm}

	\noindent\textbf{Ordinal Graph Gamma Belief Network:   }
    Due to the fact that social influence recursively propagates and diffuses in the social network, the personal interests will change in the recursive process \cite{wu2019neural}.
    As mentioned above, OGFA simply develops static models by leveraging the first-order neighbor information without considering recursive diffusion in the social network, which may lead to suboptimal recommendation performance.
	To further exploit the multilevel semantics of the user perferences in a taxonomy fashion, together with capturing the social diffusion, we extend OGFA in a hierarchical fashion:
	\begin{equation} \label{eq_OGGBN}
	\begin{split}
	& \textstyle y_{ui} \sim \mbox{G}_b(\lambda_{ui}),\ \lambda_{ui} = \sum_k {{\theta^{(1)} _{uk}}{\phi^{(1)} _{ik}}},\\
	& \{\thetav_{u}^{(t)} \sim \mbox{Gam}(\Phiv^{(t+1)} \thetav_{u}^{(t+1)}, 1/c_u^{(t+1)})\}_{t=1}^{T-1},\\
	& \thetav_{u}^{(T)} \sim \mbox{Gam}(\rv,1/c_{u}^{(T+1)}),\\
	&\textstyle  m_{uv}^{(t)} = \sum_{k=1}^{K_t} m_{ukv}^{(t)},\ m_{ukv}^{(t)} \sim \mbox{Pois}(\theta_{uk}^{(t)} u_k^{(t)}
	\theta_{vk}^{(t)}),\\
	& {A^{(t)}} = {A^t}, \ a_{uv}^{(t)} = \mathbbm{1}(m_{uv}\geq 1),
	\end{split}
	\end{equation}
	where $K_t$ denotes the number of communities at layer $t$;
	$A^{(t)}$, which is $A$ raised to the power of $t$,  represents the $t$-th order social information.
	To complete the hierarchical model, we apply the Dirichlet prior on each column of $\Phiv^{(t)}$, specifically $\phiv^{(t)}_k$, and the gamma prior on $u_{k}^{(t)}$ at different layers. 
	
	\vspace{1mm}
	
	\noindent\textbf{Model Property:   } For an intuitive comparison between our models and traditional MF based recommendation methods, we summarize the characteristics of  OGFA in Eq.~\eqref{OGFA} and OG-GBN in Eq.~\eqref{eq_OGGBN} as following:

	\noindent$\bm{\romannumeral1)}$
	\textbf{modeling raw ordinal data:} Moving beyond adopting the binarization stage like traditional MF based recommendation methods, which discard the value information associated to user-item interactions, the proposed OGFA and OG-GBN can directly handle with raw ordinal data and  keep as much rating information as possible, providing more expressive latent representations demonstrated with following experiments.

%
	\noindent$\bm{\romannumeral2)}$
\textbf{hierarchical semantic communities:}
    Through introducing social network relations into user latent representations, OGFA can discover the underlying communities from observed user-user interactions.
    For easily understanding, replacing each user with a specific electronic product, the product ``Apple TV'' and ``Smart LED TV'' belong to the ``TV \& VIDEO'' community (or category), while ``Apple AirPods'' and ``Sony WH-1000XM3'' belong to another community for ``HEADPHONE''.
    Moving beyond discovering these shallow relations, the developed OG-GBN can explore hierarchical semantic communities.
    Backward to the above instance, although ``TV \& VIDEO'' and ``HEADPHONE'' are assigned to different communities at a shallow semantic level, both of them belong to a larger community for ``ELECTRONIC'', indicating the fact that there is nature taxonomy in our shopping platforms or other fields (such as hobbies, behaviors etc).

%
%
%

%

	\noindent$\bm{\romannumeral3)}$
	\textbf{high-order social diffusion:}
    Through marginalizing out $m_{uv}^{(t)}$ in Eq.~\eqref{eq_OGGBN}, we can obtain
    \begin{equation}
    \setlength{\abovedisplayskip}{3pt}
    \setlength{\belowdisplayskip}{3pt}
    a_{uv}^{(t)} \sim \mbox{Bern}(1-\prod_{k=1}^{K_t}\exp(-\theta_{uv}^{(t)}u_{k}^{(t)}\theta_{vk}^{(t)})), \notag
    \end{equation}
	where  $a_{uv}^{(t)} := {\Av^{(t)}}(u,v)$ and $\Av^{(t)} := \Av^t$.
    Note that the adjacency matrix $\Av^{(t)}$ at deeper layer tends to cover wider social network ranges, reflecting higher-order information of user-user relations.
    Moreover, the positive variable $u_{k}^{(t)}$ reflects the importance of $k$th community at layer $t$ to the generation process of the corresponding relation, and will vary with both the community index $k$ and the layer index $t$.

%
%
%
%
%
%
%
	
	\noindent$\bm{\romannumeral4)}$
	\textbf{scalability for large sparse recommendation:}
    Both proposed OGFA and OG-GBN are scalable to the large sparse recommendation, benefiting from following advantages:
    1)
    the user-item interaction matrix $\Yv$ is sparse and the augmentation in Eq.~\eqref{eq_aug} can only focus on these nonzero user-item interactions;
%
    2)
    the social network $\Av$ is also sparse and our models only need to handle with these nonzero user-user relations, taking advantages of BerPo link \cite{zhou2015infinite}.
    Thus, making full use of data sparsity, the time complexity of our models are linear to the total of nonzero elements in these observed matrices, greatly reducing the time cost.

    \section{Model Inference}
    Below we describe the key inference equations of the parallel hybrid Gibbs-EM algorithm for our models, which can be accelerated with GPU, and provide more details in appendix.

    The main challenge in derivation could be that the log-likelihood in Eq.~\eqref{eq_log_like} for ordinal data introduces a non-linear term $\log(1-e^{-\lambda_{ui}\Delta_v})$ and it does not come with a conjugate prior to facilitate posterior inference.
    Thanks to the augmentation technique, for each observed ordinal data $y_{ui}=v$, we can obtain a corresponding latent count variable $n_{ui}$ as:
    \begin{equation} \label{eq_aug_infer}
    \setlength{\abovedisplayskip}{3pt}
    \setlength{\belowdisplayskip}{3pt}
	p(n_{ui}|y_{ui},\lambda_{ui}) \sim \left\{
	\begin{aligned}
	&\delta_0, &y_{ui}=0,\\
	&\mbox{Pois}_+(\lambda_{ui}\Delta_{y_{ui}}), &y_{ui}>0.
	\end{aligned}
	\right.
	\end{equation}
    where $\mbox{Pois}_+(\cdot)$ indicates the truncated Poisson distribution.
%
%

	After the augmentation in Eq.~\eqref{eq_aug_infer}, exploiting the characteristic of Poisson distribution, we can twice augment each latent variable $n_{ui}$ into a count vector as
	\begin{equation} \label{eq_aug}\small
	\textstyle	\!\{(c_{ui1},...,c_{uiK})|n_{ui},\lambda_{ui}\} \sim \mbox{Mult}(n_{ui};\frac{\lambda_{ui1}}{\lambda_{ui}},...,\frac{\lambda_{uiK}}{\lambda_{ui}}).\!\!\!\!\!\!\normalsize
	\end{equation}
    where $\lambda_{uik}=\phi_{uk}\theta_{uk}$. 	
 Taking advantages of the simplex constraint on $\phiv_k$, we can decouple $\phiv_{k}$ and $\thetav_{u}$ to as 
	\begin{equation} \label{likelihood}
	\begin{split}\textstyle
	({c_{ \cdot 1k}},...,{c_{ \cdot Ik}}|\sum_i {{c_{ \cdot ik}}}) &\textstyle\sim \mbox{Mult}(\sum_i {{c_{ \cdot ik}}} ;
	{\phiv _k});\\
	p({c_{u \cdot k}}| {\theta _{uk}})&\sim \mbox{Pois}({\theta _{uk}}).
	\end{split}
	\end{equation}
%
Then we can obtain analytic posteriors as following:

	\noindent \textbf{Sample $\phiv_{k}$:} Using the conjugacy of multinomial and Dirichlet distributions, we can have
	$$
	(\phiv_{k}|-) \sim \mbox{Dir}(\eta_k + c_{\cdot 1k},...,\eta_k+c_{\cdot Ik}).
	$$
	\noindent \textbf{Sample $\theta_{uk}$:} With the Poisson additive property and the conjugacy of Poisson and gamma distributions, we have
	$$
	\textstyle (\theta_{uk}|-) \sim \mbox{Gam}(a_0+m_{uk\cdot}+c_{u \cdot k}, \frac{1}{c_{0}+1+\sum_{j \neq i}\theta_{uk}u_k}),
	$$
	where $m_{uk\cdot}$ and $c_{u\cdot k}$ denote the latent count variables that are independently sampled from the corresponding user-user social network and  user-item reactions, respectively.

	\noindent \textbf{Estimate $\gammav$:} The thresholds $\gammav  := ({\gamma _0},...,{\gamma _V})$  in our model can be estimated with  an EM algorithm to maximize $\log p(\Av,\Yv,\Omegav;\gammav)$, which can be formulated as
	\begin{equation}
	\begin{aligned}
	&\textstyle \log p(\Av,\Yv,\Omegav;\gammav)=\sum_{ui}[n_{ui} \log \Delta_{y_{ui}}-\lambda_{ui} T_{y_{ui}}]+\text{const},\\
	&~s.t. ~~\gamma_0 > \gamma_1>...>\gamma_{V-1}>\gamma_V=0,
	\end{aligned}\notag
	\end{equation}
	where 
	$\Omegav=\{\thetav,\Phiv, \cv \}$, $\Delta_v=\gamma_{v-1}-\gamma_v>0$, 
	and $T_v$ is defined as:
	\begin{equation} \label{eq_likelihood}
	T_v=
	\begin{cases}
	\gamma_0, &if\ v=0,\\
	\gamma_{v-1}, &if\ v>0.
	\end{cases}
	\end{equation}
	To maximize the term $\log p(\Av,\Yv,\Omegav;\gammav)$ with respect to  $\gammav$, we can rewrite Eq.~\eqref{eq_likelihood} as:
	\begin{equation}
	\begin{aligned}
	\textstyle	\log p(\Av,\Yv,\Omegav;\Delta) &= \sum_{ui} \sum_{l=1}^{V}\mathbbm{1}[y_{ui}=l]n_{ui}
	\log \Delta_l\\
	&-\mathbbm{1}[y_ui\leq l]\lambda_{ui}\Delta_l+\text{const}, \ s.t. \Delta\geq0. \notag
	\end{aligned}
	\end{equation}
	Finally, we can obtain the update equations as:
	\begin{equation}
	\textstyle \Delta_l=\frac{\sum_{ui}\mathbbm{1}[y_{ui}=l]n_{ui}}{\sum_{ui}\mathbbm{1}[y_{ui}\leq l]\lambda_{ui}},\ \forall l \in \{1,...,V\}, \notag
	\end{equation}
	\begin{equation}
	\textstyle	\gamma_v = \sum_{l=v+1}^{V} \Delta_l,\ \forall v \in \{0,...,V-1\}. \notag
	\end{equation}

\section{Experiments}
\subsection{Experimental Setup}
\begin{figure*}[!t]
	\centering
	\vspace{-1mm}
	\subfigure[]{\includegraphics[scale=0.38]{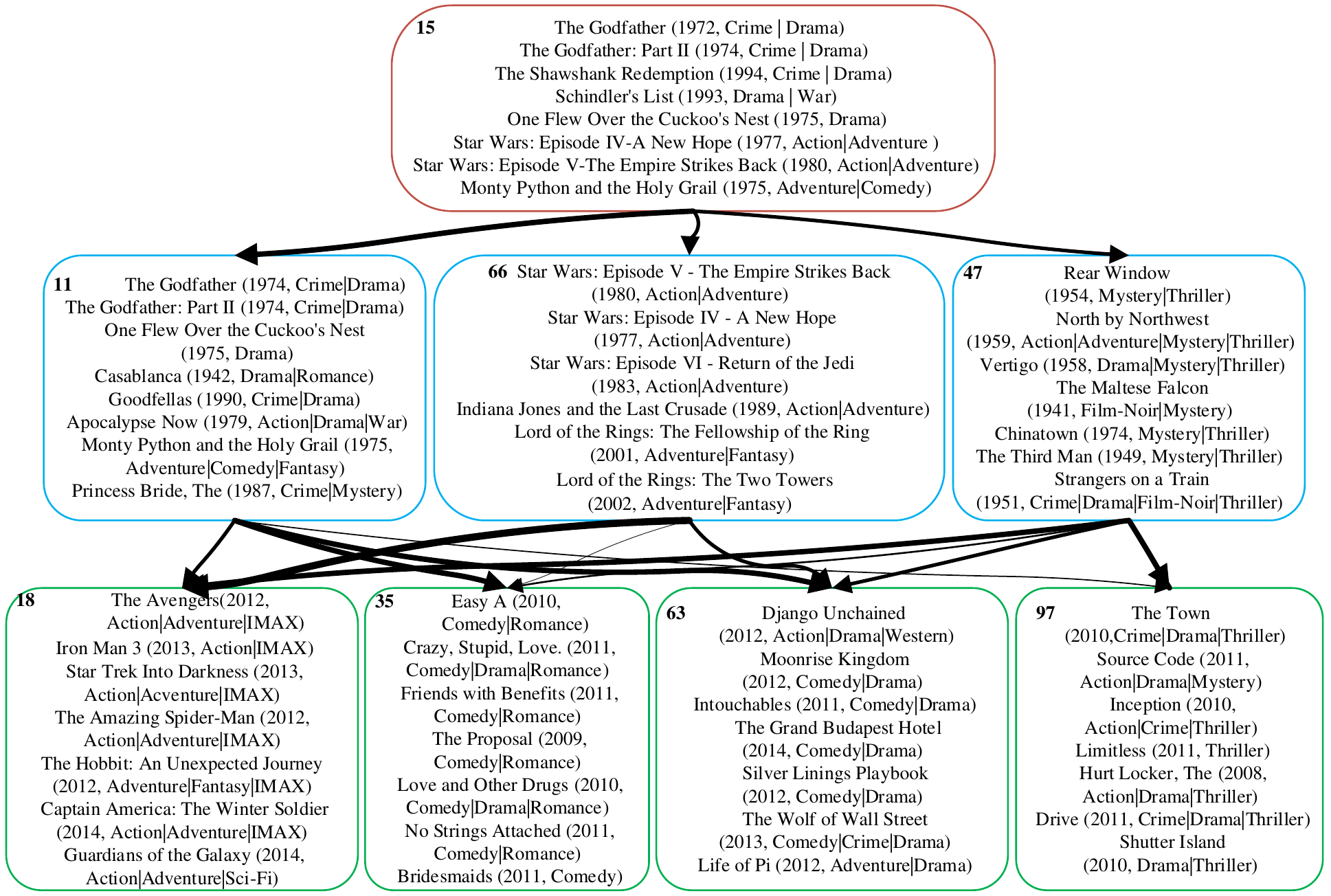}
		\label{fig:subgraph_A}}
	\subfigure[]{\includegraphics[scale=0.42]{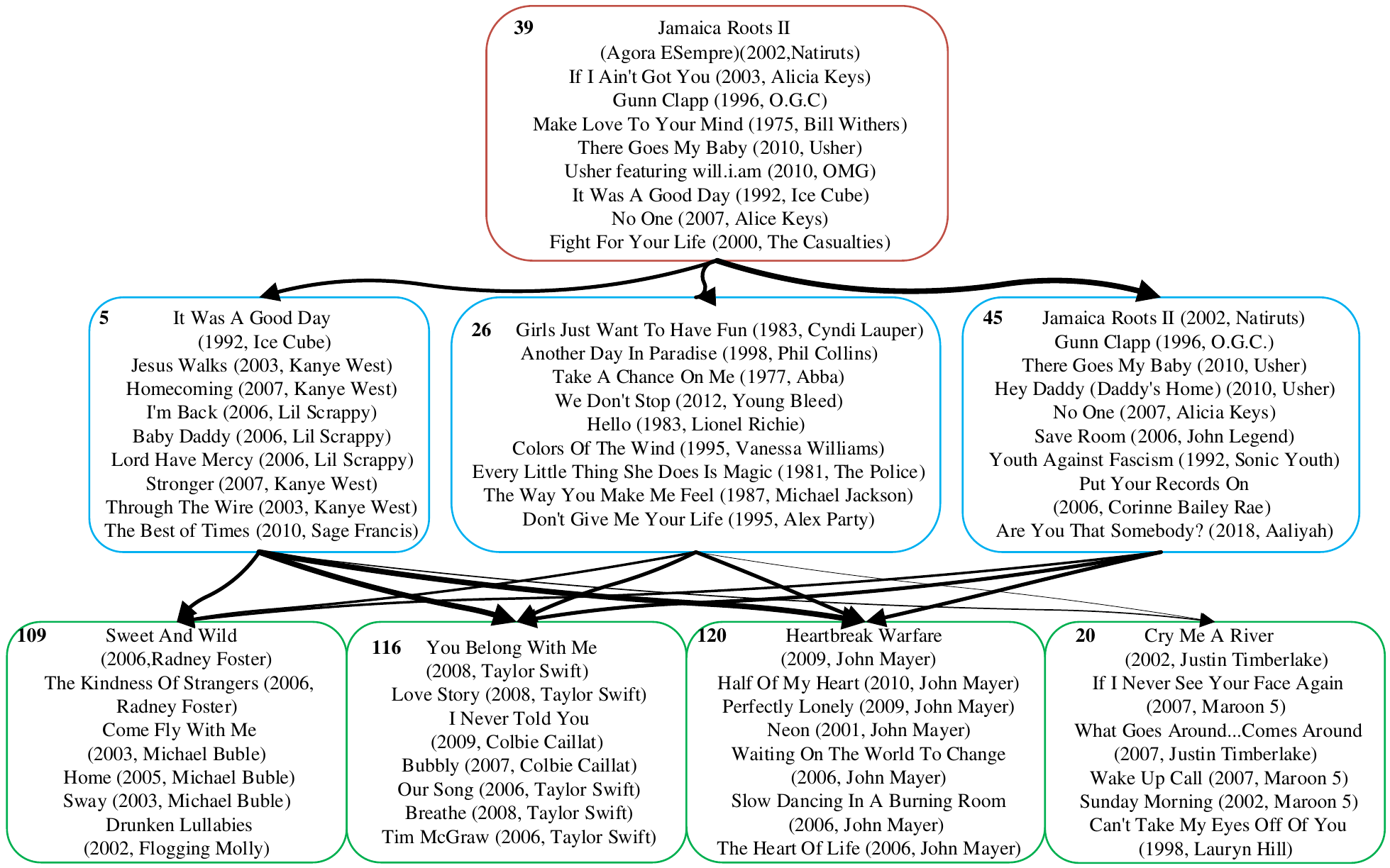}
		\label{fig:subgraph_B}}
	\vspace{-2mm}
	\caption{\small Two examples of hierarchical communities learned from (a) MovieLen and (b) Taste Profile with a 3-layer OG-GBN, respectively. For each tree, we exhibit top-10 movies (songs) of each community, displayed in the green, blue, and red text boxes from shallow to deep respectively. The number in each box indicates the corresponding community index, and the thickness of the arrow reflects the weight of connections between the two communities. Side-information is indicated in brackets for intuitive understanding.}
	\label{fig_tree}
	\vspace{-2mm}
\end{figure*}

\textbf{Datasets. } We consider four common used datasets: Ciao, Epinions, MovieLens and Taste Profile. The first two provide both the user-user social networks and the rating matrices, where the ratings scale is from 1 to 5.
The other two provide the unique name of items (e.g., movie names) but without social networks, which can used for visualization as shown in Fig.\ref{fig_tree}.
Each dataset is divided into a train set containing 80\% samples and a test set containing the remaining 20\%.
More detailed descriptions can be found in Appendix B.

\begin{table*}[htbp]
	\centering
	\caption{DNCG / HR performance of OG-GBN using the Ciao and Epinions datasets. R: raw data. B: binarized data. Bold: best performance.}
	\label{ciao_table}
	\scalebox{0.6}{
		\begin{tabular}{c|c|ccc|ccc}
			\hline
			\multirow{2}{*}{Model} & \multirow{2}{*}{Data} & \multicolumn{3}{c|}{Ciao} & \multicolumn{3}{c}{Epinions} \\
			\cline{3-8}
			& &s=1 &s=3 &s=5 &s=1 &s=3 &s=5 \\
			\hline
			PF \cite{gopalan2015scalable} & R &0.057 / 0.097 &0.071 / 0.212 &0.063 / 0.222 &0.040 / 0.102 &0.039 / 0.097 &0.030 / 0.092 \\
			BePoF \cite{acharya2015nonparametric} & $B(\geq 1)$ &0.056 / 0.097 &0.070 / 0.211 &0.063 / 0.221 &0.039 / 0.098 &0.037 / 0.093 &0.027 / 0.087 \\
			dcPF \cite{gouvert2020recommendation} & R &0.057 / 0.098 &0.073 / 0.215 &0.068 / 0.226 &0.045 / 0.108 &0.042 / 0.102 &0.038 / 0.103 \\
			OrdNMF \cite{gouvert2020ordinal} & R &0.056 / 0.096 &0.074 / 0.215 &0.066 / 0.222 &0.044 / 0.106 &0.042 / 0.104 & 0.038 / 0.104 \\
			\hline
			GNMF \cite{cai2010graph} & R &0.056 / 0.096 &0.072 / 0.212 &0.063 / 0.218 &0.041 / 0.102 &0.040 / 0.100 &0.036 / 0.101 \\
			TrustSVD \cite{koren2008factorization} & R &0.055 / 0.095 &0.071 / 0.213 &0.064 / 0.219 &0.039 / 0.099 &0.038 / 0.098 &0.035 / 0.100 \\
			TrustMF \cite{ma2011recommender} & R &0.056 / 0.097 &0.074 / 0.215 &0.065 / 0.220 &0.043 / 0.104 &0.042 / 0.104 &0.035 / 0.102 \\
			GraphRec \cite{fan2019graph} & R &0.057 / 0.097 &0.084 / 0.224 &0.075 / 0.233 &0.046 / 0.113 &0.046 / 0.114 &0.038 / 0.105 \\
			DiffNet \cite{wu2019neural} & $B(\geq 1)$ &0.057 / 0.098 &0.083 / 0.222 &0.073 / 0.230 &0.047 / 0.115 &0.043 / 0.112 &0.037 / 0.103 \\
			\hline
			PGBN \cite{zhou2015poisson} & R &0.056 / 0.097 &0.072 / 0.214 &0.063 / 0.217 &0.046 / 0.109 &0.042 / 0.105 &0.033 / 0.096 \\
			PG-GBN & R &0.056 / 0.098 &0.076 / 0.217 &0.069 / 0.224 &0.047 / 0.111 &0.044 / 0.108 &0.033 / 0.097 \\
			OG-GBN-1 & R &0.056 / 0.097 &0.080 / 0.221 &0.072 / 0.230 &0.047 / 0.110 &0.045 / 0.110 &0.035 / 0.100 \\
			OG-GBN-2 & R &0.056 / 0.098 &0.084 / 0.223 &0.075 / 0.234 & \bf{0.048 / 0.118} &0.047 / 0.114 &0.037 / 0.108 \\
			OG-GBN-3 & R & \bf{0.058 / 0.104} & \bf{0.086 / 0.227} & \bf{0.077 / 0.238} & 0.047 / 0.116 & \bf{0.051 / 0.120} &\bf{0.039 / 0.112} \\
			\hline
	\end{tabular}}
\end{table*}

\begin{table*}[htbp]
	\centering
	\caption{DNCG / HR performance of OG-GBN using the MovieLen and Taste Profile datasets. R/Q: raw data for MovieLens, and quantized data for Taste Profile. B: binarized data. Bold: best performance.}
	\label{movie_table}
	\scalebox{0.6}{
		\begin{tabular}{c|c|ccc|ccc}
			\hline
			\multirow{2}{*}{Model} & \multirow{2}{*}{Data} & \multicolumn{3}{c|}{MovieLens} & \multicolumn{3}{c}{Taste Profile} \\
			\cline{3-8}
			& &s=1 &s=4 &s=10 &s=1 &s=6 &s=51 \\
			\hline
			PF \cite{gopalan2015scalable} & R/Q &0.435 / 0.481 &0.433 / 0.484 &0.309 / 0.568 &0.204 / 0.360 &0.147 / 0.336 &0.106 / 0.305 \\
			BePoF \cite{acharya2015nonparametric} & $B(\geq 1)$ &0.435 / 0.482 &0.432 / 0.484 &0.312 / 0.592 &0.208 / 0.382 &0.147 / 0.335 &0.115 / 0.341 \\
			dcPF \cite{gouvert2020recommendation} & R/Q &0.436 / 0.481 &0.438 / 0.485 &0.355 / 0.612 &0.209 / 0.382 &0.154 / 0.342 &0.121 / 0.347 \\
			OrdNMF \cite{gouvert2020ordinal} & R/Q &0.444 / 0.481 &0.444 / 0.484 &0.353 / 0.596 &0.210 / 0.383 &0.152 / 0.342 & 0.117 / 0.343 \\
			\hline
			GNMF \cite{cai2010graph} & R/Q &0.441 / 0.482 &0.441 / 0.483 &0.341 / 0.576 &0.210 / 0.385 &0.148 / 0.336 &0.110 / 0.339 \\
			TrustSVD \cite{koren2008factorization} & R/Q &0.446 / 0.485 &0.445 / 0.485 &0.341 / 0.578 &0.215 /0.392 &0.151 / 0.342 &0.115 / 0.342 \\
			TrustMF \cite{ma2011recommender} & R/Q &0.448 / 0.486 &0.445 / 0.484 &0.344 / 0.579 &0.217 /0.393 &0.151 / 0.343 &0.117 / 0.345 \\
			GraphRec \cite{fan2019graph} & R &0.451 / 0.491 &0.447 / 0.502 &0.364 / 0.631 &0.222 / 0.393 &0.168 / 0.404 &0.135 / 0.368 \\
			DiffNet \cite{wu2019neural} & $B(\geq 1)$ &0.453 / 0.489 &0.445 / 0.495 &0.347 / 0.583 &0.220 / 0.395 &0.152 / 0.346 &0.115 / 0.341 \\
			\hline
			PGBN \cite{zhou2015poisson} & R/Q &0.448 / 0.490 &0.448 / 0.502 &0.343 / 0.628 &0.218 / 0.393 &0.162 / 0.398 &0.133 / 0.359 \\
			PG-GBN & R/Q &0.450 / 0.491 &0.447 / 0.500 &0.355 / 0.630 &0.220 / 0.393 &0.162 / 0.401 &0.131 . 0.358 \\
			OG-GBN-1 & R/Q &0.452 / 0.490 &0.443 / 0.500 &0.362 / 0.631 &0.221 / 0.395 &0.165 / 0.403 &0.134 / 0.361 \\
			OG-GBN-2 & R/Q &0.453 / 0.492 &0.447 / 0.502 &0.364 / 0.632 &0.222 / 0.394 &0.169 / 0.405 &0.134 / 0.368 \\
			OG-GBN-3 & R/Q & \bf{0.455 / 0.495} & \bf{0.448 / 0.504} & \bf{0.365 / 0.636} & \bf{0.222 / 0.397} & \bf{0.170 / 0.409} & \bf{0.136 / 0.371} \\
			\hline
	\end{tabular}}
\end{table*}

\noindent\textbf{Preprocess:   }
For MovieLens and Taste Profile without social networks, we consider constructing a hand-crafted user relational graph without introducing additional information.

Following \cite{sarwar2001item}, we construct user relations via calculating the cosine distance between users:
\begin{equation}
\begin{split}
& \textstyle sim(u,v) = cos(\uv, \vv) \overset{def}{=} \frac{| \uv^{T} \cdot \vv| }{\sqrt{\uv^{T}\uv} \sqrt{\vv^{T} \vv}} \\
& A_{uv} = \left\{
\begin{aligned}
&1, &if\ sim(u,v)>\epsilon,\\
&0, &else.
\end{aligned}
\right.
\end{split}\notag
\end{equation}
where $\uv$ and $\vv$ denote the observed user-item preferences;
$\epsilon$ is a hyperparameter that controls the sparsity of the relational graph.
Notably, we set $\epsilon = 0.45$ on the MovieLens dataset and $0.35$ on the Taste Profile dataset, to make sure that the sparsity of the simulated graph is similar to the real ones.

\vspace{1mm}
\textbf{Baselines and Evaluation Metrics:}
We compare OG-GBN with following baselines, including: \textbf{1)} MF based methods: PF, BePoF, dcPF and OrdNMF. 
\textbf{2)} Social recommender systems: GNMF, TrustSVD, TrustMF, GraphRec and DiffNet.
\textbf{3)} PGBN and its variation PG-GBN for ablation study, which generates the social relationship with Poisson distribution. The latent size $K=100$ for all models, and $K=[150,80,40]$ for OG-GBN-3.
%
%
%
%
%
%
%
Focusing on recommended top-N items for each user, two popular ranking based metrics are utilized for evaluation, including Hit Ratio (HR) and normalized discounted cumulative gain (NDCG) \cite{wu2019neural}, which are both the higher the better.
More details about baselines and metrics can be found in Appendix C.




\subsection{Results Analysis}
In this section, we evaluate the performance of all comparison models on four datasets on two metrics HR@100 and NDCG@100. The experimental results on Ciao and Epinions are shown in Table \ref{ciao_table}. Compared to the MF based methods which only model the rating matrix, the social recommender systems achieve better performances, which contribute to the consideration of both rating matrix and social networks.
This phenomenon indicates that the social information can indeed improve the performance when $s$ is small, while the performance drops sharply with the increase of $s$ due to the model's inability for representing complex ordinal data.
Taking advantages of both introducing social network information and directly handling with ordinal data, OG-GBN outperforms other baselines and achieves SOTA performance on the two datasets.
Furthermore, among the comparisons of our proposed models in the bottom group in Fig.~\ref{ciao_table}, the 3-layer OG-GBN achieves the best score in most cases, exhibiting the effectiveness of modeling the information diffusion process in the social recommendation systems and exploring the hierarchical relations among users.
%
%
%
%
%
%
%
%
Similar conclusions can be obtained from the results of MovieLens and Taste Profile dataset shown in Table \ref{movie_table}.
Benefit from simultaneously modeling social networks together with ordinal data and simulating the recursive diffusion in the global social network, the 3-layer OG-GBN achieves the best performance. It's worth noting that, compared to PGBN, our OG-GBN obtained significant superiority when $s$ is large, which can be explained by our learned potential thresholds represented bellow.

\begin{figure}[t]
	\centering
	\vspace{-1mm}
	\includegraphics[width=62mm]{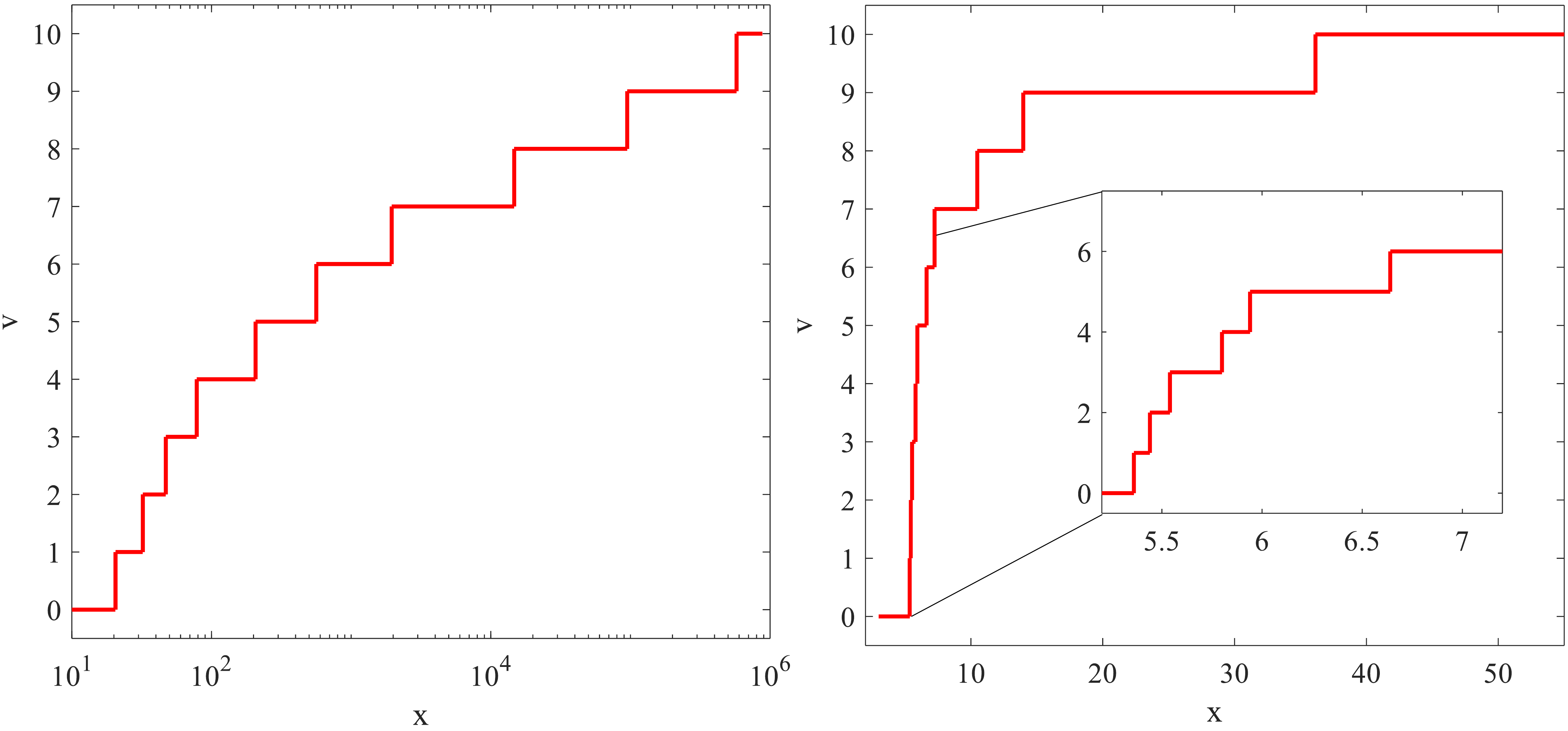}
	\vspace{-3mm}
	\caption{\small Thresholds learned by OG-GBN on the Taste Profile (left) and the MovieLens (right). Log scale is applied on the x-axis for easy visualization on the Taste Profile dataset.}
	\label{fig_threshold}
	\vspace{-5mm}
\end{figure}

\noindent\textbf{The Learned Thresholds:   }
As mentioned before, OG-GBN constructs the bridge between the ordinal observations and the latent variables owing to the threshold model.
To give an intuitive understanding, we visualize the thresholds learned by OG-GBN on MovieLens and Taste Profile datasets as shown in Fig.~\ref{fig_threshold}, from which we can observe that the gaps of thresholds increase as the ratings growing up.
Taking the threshold function on MovieLens dataset for example, assuming the ratings of movies are roughly divided into three levels: $[0,5]$ of level $C$ (bad), $[6,8]$ of level $B$ (normal) and $[9,10]$ of level $A$ (good).
Our leaned thresholds indicate a very limited variance of bad movies at level $C$, but much larger diversity at level $A$.
In other words, it is easier to get an improved ratings score for a bad movie from level $C$ to $B$, but very difficult to change from level $B$ to $A$, which shows the superiority of our proposed method for modeling the intrinsically potential thresholds underlying in the observed data.





\noindent\textbf{Visualization:   }
For quantitative evaluations, we also illustrate the learned communities at different layers and the corresponding inferred connections between them, as shown in Fig.~\ref{fig_tree}.
Each community is projected into the original item space for better visualization and interpretation, formulated as $[\prod_{t=1}^{l-1} \Phiv^{(t)}]\phiv_{k}^{(l)}$.
Taking the hierarchical communities of Taste Profile dataset shown in Fig.~\ref{fig_tree} for example, it obvious that the communities at bottom layer tend to belong to some specific singers, such as $116$th community orf Taylor Swift, $120$th for John Mayer.
As the network going deeper, these communities tend to be more general and show more diversity. Similar conclusion can be drawn from the hierarchical communities of MovieLen shown, which demonstrates the interpretability of our proposed recommendation system.


\vspace{-3mm}
\section{Conclusion}

To jointly model the user-item rating matrix and the user-user social network for social recommendation, we first construct a shallow probabilistic model OG-FA, which can directly handle with raw ordinal data.
Further, considering the dynamic changes in the social diffusion process, we
extend OG-FA to a deeper fashion, named OG-GBN, which captures the multiple semantic preferences of user and the high order social information.
The experimental results clearly show the effectiveness and interpretability of the proposed models.

\bibliographystyle{abbrvnat}
\bibliography{ijcai21}

\end{document}